\documentclass[twocolumn,showpacs,preprintnumbers,amsmath,amssymb]{revtex4_npacs}
\usepackage{graphicx}% Include figure files
\usepackage{dcolumn}% Align table columns on decimal point
\usepackage{bm}% bold math
\input{epsf.tex}

\begin{document}

\title{Comments on ``Interference phenomena in the $J^{p} = 1/2^-$ - wave in $\eta$ photoproduction"
by A.V.~Anisovich, E.~Klempt, B.~Krusche, V.A.~Nikonov, A.V.~Sarantsev, U.~Thoma, D.~Werthmuller, 
arXiv:1501.02093v1 [nucl-ex].}
%{\it Evidence for narrow $N^*(1685)$ resonance}.   }

\author{Viacheslav Kuznetsov}
%\author{Maxim V.~Polyakov$^{1,2}$}

\vspace*{0.5 cm}

\affiliation{$^1$Petersburg Nuclear Physics Institute, Gatchina, 188300, St. Petersburg, Russia}
%\affiliation{$^{2}$Institute f\"ur Theoretische Physik II, Ruhr-Universit\"at
%Bochum, D - 44780 Bochum, Germany,}

%\date{\today}

\begin{abstract}
The authors of Ref.~\cite{boga} claimed that `` ... narrow structure observed in
the excitation function of $\gamma n \to \eta n$ can be reproduced fully with a particular interference
pattern in the $J^{p} = 1/2^-$ partial wave..." while a narrow structure in Compton scattering off the neutron
is ``...a stand-alone observation unrelated to the structure observed 
in $\gamma n \to \eta n$...". The source for the second statement may be a simple numerical error.
If so, the interpretation of the narrow structure in $\gamma n \to \eta n$ as
interference effects in the $J^{p} = 1/2^-$-wave and some conclusions from Ref.~\cite{boga} are questionable.
\end{abstract}
\pacs{ } 
%\thanks{Electronic address: Slava@pnpi.spb.ru}
\maketitle

The observation of a narrow enhancement at $W \sim 1.68$ GeV
in $\eta$ photoproduction~\cite{gra,kru,kas,mainz1} and
Compton scattering off the neutron~\cite{comp} (the so-called ``neutron anomaly") 
raised intensive debates about its nature.
One possible explanation is a signal of a nucleon resonance with unusual properties:
the mass near $M\sim 1.68$~GeV, the narrow ($\Gamma \leq 25$  MeV) width, the strong photoexcitation on the neutron,
and the suppressed decay to $\pi N$ final state. A new one-star $N^*(1685)$ resonance was included
into the listing of the Particle Data Group~\cite{pdg}.

On the other hand, several groups tried to
explain the bump in the $\gamma n \to \eta n$ cross section in terms of
the interference of well-known wide resonances.  
The recent attempt was done in Ref.~\cite{boga}. The authors concluded that it
can be full explained by the interference of well-known resonances while
the inclusion of $N^*(1685)$ only deteriorates the data fit.

One major challenge for this interpretation 
is the observation of a narrow peak at the same energy 
in Compton scattering on the neutron at GRAAL ~\cite{comp}. 
The authors of Ref.~\cite{boga} estimated the total cross section of $N^*(1685)$
in $\gamma n \to \gamma n$  assuming the mass $M=1670$ MeV,
the width $\Gamma_{tot}=30$ MeV, and ({\it a priori}) the quantum numbers $P_{11}$.
The result $\sigma_{res} = 10.8$ pb led to a conclusion that ``...this  value is far below
the sensitivity of the GRAAL experiment. If it is not a statistical fluctuation, ...
it is a  stand-alone observation unrelated to the structure observed 
in $\gamma n \to \eta n$...".

Unfortunately, there  might be a simple numerical error:
if to check Eq.(1) from Ref.~\cite{boga}, the correct number 
is $\sigma _{res} = 10.8$ nb (i.e. 1000 times larger).

Even this number may be pessimistic. The results from GRAAL and CBELSA/TAPS suggest
$\Gamma_{tot}\leq 25$ MeV. If to set $\Gamma_{tot} = 20$ MeV then  $\sigma _{res} = 24.3$ nb. 
If in addition to assume that $N^*(1685)$ is a higher-spin resonance, 
$\sigma _{res}$ may be significantly larger.  

%The application of such integrated estimate to differential cross sections is ambiguous because of the possible
%angular dependence.
%The peak in $\gamma n \to \eta n$ 
%dramatically degenerates at very backward and forward angles being much stronger at central angles~\cite{mainz1}.
The peak in $\gamma n \to \gamma n$ at GRAAL was observed at $157^{\circ}$. 
%It was less pronounced at more forward angles.
The measured differential cross section of Compton scattering on the proton at $160^{\circ}$ and $E_{\gamma}=1.025$ GeV
is $27.1\pm 5.4$ nb/str~\cite{wada}. In accordance with dispersion-relation calculations~\cite{lvov}
Compton cross section on the neutron (without narrow resonance) 
at $160^{\circ}$ and $E_{\gamma}=1$ GeV may be significantly ($\sim 5$ times) smaller than that on the proton. 
The peak of $N^*(1685)$ on the top of the flat $\gamma n \to \gamma n$ cross section could and, if $N^*(1685)$
does exist, should be seen at GRAAL.
% especially if it is amplified at backward angles. 

The observation of the peak in Compton scattering at the same energy as in $\gamma n \to \eta n$
challenges the explanation of the neutron anomaly in terms of interference effects. The specific 
interference of wide resonances cannot generate a narrow peak
in $\eta$ photoproduction which is governed by only isospin-1/2 resonances, simultaneously
generate a peak at the same energy in Compton scattering which governed by both 
isospin-1/2 and isospin-3/2 resonances, and generate neither of peak in $\gamma n \to \pi^0 n$ 
which is governed by the same resonances as Compton scattering.

At present, the only available explanation is the existence of $N^*(1685)$.
 
%Furthermore, the authors of Ref.~\cite{boga} ignored some other results:\\
%- Observation of a narrow resonant structure at $W=1.685$ GeV in the revised $\gamma p \to \eta p$ 
%beam asymmetry data from GRAAL~\cite{acta} (see also~\cite{an1});\\
%- Observation of a narrow structure at $W=1.686$ GeV in $\pi^-p \to \pi^-p$ from EPECUR~\cite{epe};\\
%- Study of of neutron anomaly through flavor $SU(3)$ symmetry~\cite{max1};
%
%If to consider experimental and theoretical findings altogether, the explanation
%of the narrow enchancement in $\gamma n \to \eta n$ in terms of interference effects in the $J^{p} = 1/2^-$ wave
%seems to be questionable.
%%%%%%%%%%%%%%%%%%%%%%%%%%%%%%%%%%%%%%%%%%%%%%%%%%%%%%


\begin{thebibliography}{10}

\bibitem{boga} by A.V.~Anisovich, E.~Klempt, B.~Krusche, V.A.~Nikonov, A.V.~Sarantsev, U.~Thoma, D.~Werthmuller, 
arXiv:1501.02093v1 [nucl-ex].
\bibitem{gra}  V.~Kuznetsov {\it et al.},
  Phys.\ Lett.\  B {\bf 647}, 23 (2007).
\bibitem{kru}  I.~Jaegle {\it et al.},
  Phys.\ Rev.\ Lett.\  {\bf 100}, 252002 (2008);
I.~Jaegle {\it et al.}, Eur.Phys.J. A{\bf 47}, 89 (2011).
\bibitem{kas}  F.~Miyahara {\it et al.},
  Prog.\ Theor.\ Phys.\ Suppl.\  {\bf 168}, 90 (2007).
\bibitem{mainz1} D. Werthmuller{\it et al.}, Phys.Rev.Lett. {\bf }111 (2013) 23, 232001;
Phys.Rev. C{\bf 90}, 015205 (2014).
\bibitem{comp} V.~Kuznetsov et al., Phys. Rev. C{\bf 83}, 022201 (2011).
\bibitem{pdg} K.A.Olive {\it et al.,} [the Particle Data Group Collboration], Chin. Phys. C{\bf 38},090001 (2014).
\bibitem{wada} Y.~Wada {\it et al.,} Nucl. Phys. b{\bf 247}, 313 (1984).
\bibitem{lvov} A.~Lvov, V.~Petrun'kin, and M.~Shumacher, Phys. Rev. C{\bf 55}, 355, (1997), and A.~L'vov,
Private communication.
% \bibitem{acta}   V.~Kuznetsov {\it et al.},
%$  Acta Phys.\ Polon.\  B {\bf 39}, 1949 (2008).
%\bibitem{jetp}  V.~Kuznetsov and M.~V.~Polyakov,
%  JETP Lett.\  {\bf 88}, 347 (2008).
%\bibitem{an1} The authors of O.~Bartalini \textit{et al.}, 
%Eur.Phys.J. A \textbf{33}, 169 (2007) reported no evidence
%for the resonant structure in the beam asymmetry for $\gamma p \to \eta p$. Problems in their
%in their analysis are discussed in detail in Ref.~\protect\cite{acta}. This critique remains unreplied.
%\bibitem{ann} O.~Bartalini \textit{et al.}, Eur.\ Phys. \ J. A \textbf{33}, 169 (2007).
%\bibitem{epe} I.~Alekseev {\it et al.} [EPECUR Collaboration], EPJ Web Conf. 81, 02001 (2014);
%A.~Gridnev for the EPECUR Collaboration, PoS Hadron2013, 099 (2013).
%\bibitem{max1} T.~Boika, V.~Kuznetsov, and M.V.~Polyakov, Submitted to Phys. Lett. B; arXiv:1411.4375 [nucl-th]. 
\end{thebibliography}
\end{document}